\documentclass[10pt]{article}
\usepackage{amsmath,amssymb,amsthm,amsfonts,color,colordvi}

\newcommand{\beq}{\begin{equation}}
\newcommand{\eeq}{\end{equation}}
\newcommand{\bsplit}{\begin{split}}
\newcommand{\esplit}{\end{split}}

\newcommand{\ket}[1]{\vert#1\rangle}

\title{Simulatings POVMs on EPR pairs with six bits of expected communication}
\author{Andr\'{e} Allan M\'{e}thot\thanks{Laboratoire d'infomatique th\'eorique et quantique, D\'epartement d'informatique et de recherche op\'erationnelle, Universit\'e de Montr\'eal, CP 6128 succ Centre-Ville, Montr\'eal QC H3C 3J7, Canada, methotan@iro.umontreal.ca}
\vspace{0.75 cm}\\
Laboratoire d'infomatique th\'eorique et quantique\\
D\'epartement d'informatique et de recherche op\'erationnelle\\
Universit\'e de Montr\'eal
}
\date{\today}

\begin{document}

\maketitle
\begin{abstract}
We present a classical protocol for simulating correlations obtained by bipartite POVMs on an EPR pair. The protocol uses shared random variables (also known as local hidden variables) augmented by six bits of expected communication.
\end{abstract}
\vspace{1 cm}

Entanglement simulation was first introduced by Maudlin in a 1992 paper published in a philosophical journal \cite{maudlin92} and was revived independently by Brassard, Cleve and Tapp in 1999 \cite{bct99}. The objective was to quantify the non-locality of EPR pairs \cite{bell64} in terms of the amount of communication necessary to simulate the correlations obtained by bipartite measurement of an EPR pair. ``The key to understanding violations of Bell's inequality is not operator algebras but information transmission \cite{maudlin92}.'' This approach increases our understanding of the relationships between classical information and quantum information. It also helps us gauge the amount of information hidden in the EPR pair itself or, in some sense, the amount of information that must be space-like transmitted, in a local hidden variable model, in order for nature to account for the Bell inequalities.\\

In this scenario, Alice and Bob try to output $a$ and $b$ respectively, through a classical protocol, with the same probability distribution, hence correlations, as if they shared an EPR pair and each measured his or her half of the pair according to a given random von Neumann measurement. In \cite{bct99}, a protocol using an infinite amount of random shared variables and eight bits of communication in the worst case was given. Independently, Steiner \cite{steiner99} presented a protocol using an infinite amount of random shared variables and 1.48 bits of \emph{expected} communication for the simulation of von Neumann measurements in the \emph{real plane}. Surprisingly, Maudlin \cite{maudlin92} already had a protocol using random shared variables and 1.17 bits of expected communication to also simulate von Neumann measurements in the real plane. This later result was improved by Cerf, Gisin and Massar \cite{cgm99} to 1.19 bits of expected communication, still with an infinite amount of shared variables, for \emph{arbitrary} von Neumann measurements. They were also able to generalize their protocol to simulate POVMs with 6.38 bits of communication. Although both type of simulations, worst-case and expected, used shared random variables so far, it was shown by Massar {\it et al.} \cite{mbcc00} that only protocols of the worst-case communication type required shared randomness. In fact, when considering the simualtion of quantum entanglement with a protocol that uses a bounded amount of communication, an infinite amount of shared variables is needed. In \cite{mbcc00}, a protocol to simulate POVMs with 20 bits of expected communication, without \emph{any} shared randomness, was detailed. Subsequent refinements have been made on the result of Brassard, Cleve and Tapp. Csirik \cite{csirik02} proposed a protocol using only six bits of communication and recently Toner and Bacon \cite{tb03} presented a protocol using only \emph{one} bit of communication. For a survey, one can be refered to \cite{brassard01}.\\

In this paper, we present a protocol, based on \cite{tb03} and \cite{cgm99}, to \emph{simulate arbitrary POVMs} on an EPR pair \emph{using six bits of expected communication}. First, we decribe what a POVM is, what is the probability distribution of the outputs if we actually make POVMs on a $\ket{\phi^+}=\frac{1}{\sqrt{2}}(\ket{00}+\ket{11})$ state and display some tools that we will need for the protocol. Then a description of the protocol is given followed by an analysis.\\

A POVM is a family of matrices $\{B_i\}$ such that $\sum A_i^{\dagger}A_i=\sum B_i =\text{I}$, where $B_i$ is called a \emph{POVM element}. On qubits, the POVM elements can be expressed, without lost of generality, as the linear sum of one-dimensional projectors \cite{cgm99}. Thus one can find vectors on the Bloch sphere $\vec{b}_i$ such that $B_i=(\vert\vec{b}_i\vert \text{I}+\vec{b}_i\cdot\vec{\sigma})/2$, where $\vec{\sigma}$ are the Pauli matricies and with the completeness conditions $\sum\vert\vec{b}_i\vert =2$ and $\sum\vec{b}_i=0$. Let's assume that Alice and Bob share a $\ket{\phi^+}$ state. Alice receives the description of a POVM $\{A_i\}$ and Bob receives the description of a POVM $\{B_j\}$. If they each measure their half of the state according to their description of the POVM, Alice will produce $a=i$ with probability $\Pr[a=i]=\vert\vec{a}_i\vert/2$, Bob will produce $b=j$ with probability $\Pr[b=j]=\vert\vec{b}_j\vert/2$ and the joint probability will be $\Pr[a=i,b=j]=(\vert\vec{a}_i\vert\vert\vec{b}_j\vert+\vec{a}_i\cdot\vec{b}_i)/4$.\\

Now let us turn our attention to the presentation of the protocol. The classical protocol (with six bits of expected communication) to simulate an arbitrary bipartite POVM on a $\ket{\phi^+}$ can be described as follows:

\begin{itemize}
\item Alice and Bob share two random unit vectors $\vec{v}_1, \vec{v}_2\in \mathbb{R}^3$.
\item Alice and Bob are given a description of their POVM $\{A_i\}$ and $\{B_i\}$ respectively.
\item Alice choses the $i^{\text{th}}$ output of her POVM according to the probability distribution $\Pr[a=i]=\vert\vec{a}_i\vert /2$.
\item Alice sends $c=\Theta (-\vec{a}_i\cdot\vec{v}_1)$ and $d=\Theta (-\vec{a}_i\cdot\vec{v}_2)$.
\item Bob choses the $j^{\text{th}}$ output of his POVM according to the probability distribution $\Pr[b=j]=\vert\vec{b}_j\vert /2$.
\item Bob checks if $\vec{b}_j\cdot((-1)^c\vec{v}_1+(-1)^d\vec{v}_2)<0$, if so he sends $0$ to Alice and they start over with a fresh set of random variables.
\item Otherwise, Bob sends $1$ to Alice and they produce their output $i$ et $j$ respectively.
\end{itemize}

The analysis of the protocol is quite simple. The probability of Alice obtaining the POVM outcome $i$ is, as stated, $\Pr[a=i]=\vert\vec{a}_i\vert /2$. As for Bob's marginal probability distribution, the vector $(-1)^c\vec{v}_1+(-1)^d\vec{v}_2$ can be considered as a vector pointing in a random direction. Therefore, Bob has probability $1/2$ of rejecting $\vec{b}_j$. Since each time around the probabilities are independent, Bob's marginal probability is $Pr[b=j]=\vert\vec{b}_j\vert /2$. The joint probability distribution, the calculation is a bit tricky but is still straightfoward:
\nopagebreak
\begin{equation}
\label{analyse}
\begin{split}
\Pr[a=i,b=j]= & \frac{\vert\vec{a}_i\vert\vert\vec{b}_j\vert}{4(4\pi)^2}\iint d\vec{v}_1d\vec{v}_2\ \Theta(\vec{b}_j\cdot ((-1)^c\vec{v}_1+(-1)^d\vec{v}_2))\\
= & \frac{\vert\vec{a}_i\vert\vert\vec{b}_j\vert}{4(4\pi)^2}\iint d\vec{v}_1d\vec{v}_2\ \Theta(\vec{b}_j\cdot (sgn(\vec{a}_i\cdot\vec{v}_1)\vec{v}_1+sgn(\vec{a}_i\cdot\vec{v}_2)\vec{v}_2))\\
= & \frac{\vert\vec{a}_i\vert\vert\vec{b}_j\vert}{4(4\pi)^2}\iint d\vec{v}_1d\vec{v}_2\ \Theta(sgn(\vec{a}_i\cdot\vec{v}_1)\vec{b}_j\cdot (\vec{v}_1+sgn(\vec{a}_i\cdot\vec{v}_1)sgn(\vec{a}_i\cdot\vec{v}_2)\vec{v}_2))\\
= & \frac{\vert\vec{a}_i\vert\vert\vec{b}_j\vert}{4(4\pi)^2}\iint d\vec{v}_1d\vec{v}_2\ \left(\frac{1+sgn(\vec{a}_i\cdot\vec{v}_1)sgn(\vec{b}_j\cdot (\vec{v}_1+sgn(\vec{a}_i\cdot\vec{v}_1)sgn(\vec{a}_i\cdot\vec{v}_2)\vec{v}_2))}{2}\right)\\
= & \frac{\vert\vec{a}_i\vert\vert\vec{b}_j\vert}{8}+\frac{\vert\vec{a}_i\vert\vert\vec{b}_j\vert}{8(4\pi)^2}\iint d\vec{v}_1d\vec{v}_2\ sgn(\vec{a}_i\cdot\vec{v}_1)sgn(\vec{b}_j\cdot (\vec{v}_1+sgn(\vec{a}_i\cdot\vec{v}_1)sgn(\vec{a}_i\cdot\vec{v}_2)\vec{v}_2))\\
= & \frac{\vert\vec{a}_i\vert\vert\vec{b}_j\vert+\vec{a}_i\cdot\vec{b}_j}{8}.
\end{split}
\end{equation}

For more details on the integration, see \cite{tb03}. As for the $1/2$ factor in~\eqref{analyse}, it is removed by renormalization (which is allowed since all instances are independent from one another) \cite{cgm99}. To realize the protocple Alice must send two bits to Bob and Bob one bit to Alice. Since each round is independent of the preceding ones and each has probability of $1/2$ of ending the protocol, the protocol takes an average of two rounds, hence $2(2+1)=6$ bits of communication. If we are allowed block-coding, the communication can be lowered. Alice still sends $c$ to Bob, but she can send $d'=\Theta(\vec{a}_i\cdot\vec{v}_1)\oplus\Theta(\vec{a}_i\cdot\vec{v}_2)$ from which $d$ can be easily recovered. From \cite{tb03}, we know that $d'$ can be compressed to an average of $0.85$ bits \cite{tb03}. The communication becomes $2(1+0.85+1)=5.7$ bits.\\

Although this result is a small improvement and a simplification of the result in \cite{cgm99}, much is left to do. We do not know any way to simulate POVMs with worst-case communication or even if it is possible to do so. While von Neumann measurements can be done with worst-case communcation, it is not clear that they can be done for POVMs. Von Neumman measurements have only two possible outcomes while POVMs have an unbounded number of outcomes. Can a protocol with a bounded amount of communication choose correctly between an unbounded number of outcomes? Can we generalize these types of protocols to simulate arbitrary measurements on $n$ EPR pairs? GHZ and other entangled states?\\

We would like to thank Ben Toner and Gilles Brassard for very helpful discussions and Anne Broadbent for helpful comments on the manuscript.

\bibliographystyle{acm}
\bibliography{bibliopovm}
\end{document}